\documentclass[12pt,preprint]{aastex}

\shorttitle{Radial Modes of Neutron Stars with a Quark Core}
\shortauthors{P. K. Sahu et al.}

\begin{document}

\title {Radial Modes of Neutron Stars with a Quark Core}

\author{P.K. Sahu, G.F. Burgio and M. Baldo} 
\affil{Istituto Nazionale di Fisica Nucleare, Sezione di Catania \\
Corso Italia 57, I-95129 Catania, Italy}

\begin{abstract}
We make a first calculation of eigenfrequencies of radial pulsations of 
neutron stars with quark cores in a general 
relativistic formalism given by Chandrasekhar.
The equations of state (EOS) used to estimate such eigenfrequencies have 
been derived by taking proper care of the hadron-quark phase transition.
The hadronic EOS's have been obtained in the framework of the 
Brueckner-Hartree-Fock 
and relativistic mean field theories, whereas the quark EOS has been derived 
within the MIT bag model.
We find that the periods of oscillations of neutron stars 
with a quark core show a kink, which is associated with the presence 
of a mixed phase region.
Also, oscillation periods show significant differences 
between ordinary neutron stars and neutron stars 
with dynamical quark phases.
\end{abstract}


Nuclear matter at sufficiently high density and temperature is expected
to undergo a phase transition to a quark-gluon plasma. 
The indirect evidences from heavy-ion experiments at CERN 
\citep{heinz00,jacobs00} and 
more recently at RHIC \citep{blaizot01} assume to confirm the formation 
of a quark-gluon plasma.
Such a phase transition might occur inside neutron stars,
because these are cold and very compact  astrophysical objects. 
It is therefore very interesting to study the effects of possible
phase transitions in neutron stars observable like
maximum gravitational masses, radii, oscillation frequencies, etc.
\par
In the present letter, we analyze the consequences of a hadron-quark
phase transition on the periods of radial oscillations in 
neutron stars. In fact, more than three decades ago, 
Cameron (1965) made a suggestion that vibration of neutron stars 
could excite motions that can have interesting astrophysical implications. 
There are several investigations of vibrating neutron stars and the simple 
dimensional analysis suggest that the period of fundamental mode would be
the order of milliseconds.
More than two decades later, Cutler et al. (1990) concluded that 
neutron stars of about one solar mass and radius about 10 km give periods 
(3--5) ms, and these turned out to be relatively insensitive to the exact 
value of central 
density. Also, Datta et al. (1992) tried  to calculate the oscillation 
periods for strange quark stars and they found it to lie in a range of 
values 0.06--0.3 ms. 
These values were not substantially different from the ones of 
conventional neutron stars 
characterized by the periods $\sim$ 0.3 ms for primary mode and 
$\le$ 0.2 ms for higher 
modes. Since the neutron stars were assumed to be composed of only hadron 
matter, the results may be different if a core made 
of quark matter is present. Of course, the  results strictly depend on the 
construction of the equation of state along with some constraints on the  
parameters in both hadron and quark phases. Here we are mainly searching
for possible signals from the onset of the quark phase, therefore the 
details of the EOS should not be relevant.
\par
In this letter, we adopt the equations of state for neutron stars 
with a quark core
as developed very recently by Burgio et al. (2001), and then we estimate 
the period of radial 
oscillations by using  pulsation equations of a nonrotating star in 
general relativistic formalism given by Chandrasekhar (1964).
\par
The equation of state used here may be divided into three components. 
The equations of state used in the hadron sector have been 
derived in the non-relativistic Brueckner-Hartree-Fock (BHF) 
and the relativistic mean field (RMF) approaches.
In the BHF method \citep{baldo99}, the
Brueckner-Bethe-Goldstone formalism has been used 
with realistic Paris two-body \citep{lacombe80} and Urbana 
\citep{carlson83,schiavilla86} three-body forces,
to ensure the correct reproduction of nuclear matter saturation properties. 
This procedure has been extended to asymmetric nuclear 
matter \citep{baldo98,baldo00,vidana00} including
hyperons by implementing hyperon-nucleon potentials that are fitted with 
existing scattering data.
The RMF theory has been derived \citep{serot86} from the many-body Lagrangian 
density in the mean field approximation. 
The parameters are fitted in such a way that it reproduces the correct values
of nuclear matter properties at the saturation point \citep{ghosh95,sahu00}.
Also hyperons are included in RMF at the mean field level for asymmetric 
nuclear matter with symmetry energy around 30 MeV at saturation, same as that
of BHF theory.
The nuclear incompressibility is around 260 MeV for BHF theory and is taken 
to be same in RMF theory for the sake of compatibility.
\par
The quark matter equation of state has a big uncertainty due to models 
dependence and its 
parameters. In the literature, there are models \citep{schertler98} associated 
with different ad hoc parameters in quark masses and bag constant.
We adopted here the simple MIT bag model \citep{chodos74} both with density 
dependent and independent bag constant.
The density dependent bag constant is fixed according to the hypothesis of 
a constant energy density along the transition line, compatible with the
CERN data. Several parametrizations have been considered in recent calculations
\citep{burgio01}. For a Wood-Saxon parametrization of the 
density dependent bag constant, the onset of quark phase in 
neutron stars takes place at about two times the saturation density,
while for Gaussian like parametrizations the onset occurs at lower density.
To be specific we choose the Wood-Saxon parametrization, but the results 
are expected to be similar for the other parametrizations.
\par
Once the equations of state both in the hadron and the quark sector are 
well established,
we can construct the mixed phase by assuming a first order phase transition. 
As pointed out by Glendenning (1992), the first order phase transition 
in neutron stars
differs from the one in ordinary matter, because of two conserved charges, i.e.
the baryon charge and the global electrical charge. As a consequence, 
the pressure in the mixed phase varies continuously with the baryon density
and is not a constant. The proportion of the hadron and quark component in 
the mixed phase is then calculated imposing the mechanical and chemical 
equilibrium, supplemented by the condition of global charge neutrality 
\citep{schertler98,burgio01}.
Finally we can construct the total equation of state, which spans  
from hadron to mixed and quark phases. This is the main ingredient needed 
to calculate the frequencies of radial modes in neutron stars.
It has to be noticed that this equation of state, according to the
complete Glendenning' s construction, includes the mixed phase and there is
no sharp surface, at a given density (and radius), separating the hadron 
and the quark phases. This is at variance with the calculations of
Haensel et al. (1989) and represents one of the novelty of our 
calculations. See also Haensel et al. (1990).
In agreement with this physical construction, during 
the star oscillations, the transition from one phase to the other cannot 
be limited by diffusion. Since the time scale of weak processes are surely 
much shorter than the period, matter will remain in beta 
equilibrium and the calculated equation of state is the relevant one
for the study of the density oscillations.
\par

The equation for infinitesimal radial pulsations of a nonrotating star 
was given by Chandrasekhar (1964)  and, 
in general relativity formalism,  has the following form:

\begin{equation}
X \frac {d^2\xi}{dr^2} + Y \frac{d\xi}{dr} + Z \xi = \sigma^2\xi. 
                                                   \label{puls}
\end{equation}
Here $\xi$(r) is the Lagrangian fluid displacement and
$c\sigma$ is the characteristic eigenfrequency ($c$ is the speed
of light). 
The quantities $X$, $Y$, $Z$ depend on the equilibrium profiles 
of the pressure $p$ and density $\rho$ of the star and are represented by

\begin{eqnarray}
X &=& \frac{- e^{-\lambda} e^{\nu}}{p + \rho c^2} \Gamma p
                                 \label{f} ,\\
Y&=&\frac{- e^{-\lambda} e^{\nu}}{p + \rho c^2}   \Bigl\{\Gamma p \Big(\frac {1}{2}
       \frac{d\nu}{dr} + \frac{1}{2} \frac{d\lambda}{dr} +
       \frac{2}{r}\Big)  \nonumber \\
      &+& p \frac{d\Gamma}{dr} + \Gamma \frac{dp}{dr}\Bigr\},\\
Z &=&\frac {e^{-\lambda}e^{\nu}}{p + \rho c^2} \Bigl\{ \frac{4}{r}
       \frac{dp}{dr} - \frac{(dp/dr)^2}{p + \rho c^2} - A \Bigr\} \nonumber \\
     &+& \frac{8\pi G}{c^4} e^{\nu} p
\end{eqnarray}
$\Gamma$ is the adiabatic index defined as

\begin{equation}
\Gamma = (1 + \rho c^2/p) \frac{dp}{d(\rho c^2)} ,
\end{equation}

\noindent and

\begin{eqnarray}
A &=&\frac {d\lambda}{dr} \frac{\Gamma p}{r} + \frac
         {2p}{r} \frac{d\Gamma}{dr} + \frac {2\Gamma}{r} \frac{dp}{dr} -
         \frac{2\Gamma p}{r^2}\nonumber\\
        &-&\frac{1}{4} \frac{d\nu}{dr} \Big( \frac{d\lambda}{dr} \Gamma p +
         2p \frac{d\Gamma}{dr} + 2\Gamma \frac{dp}{dr} - \frac{8\Gamma
         p}{r}\Big)\nonumber\\ 
       & - & \frac{1}{2} \Gamma p
         \Big(\frac{d\nu}{dr}\Big)^2 - \frac{1}{2}
         \Gamma p \frac{d^2\nu}{dr^2}.   \label{a}
\end{eqnarray}
To solve the pulsations equation (\ref{puls}), the boundary conditions are
\begin{eqnarray}
\xi (r = 0)&=&0,  \label{xi0} \\ 
\delta p (r = R) &=& -\xi \frac{dp}{dr} - \Gamma p \frac
                   {e^{\nu/2}}{r^2} \frac{\partial}{\partial r}
                   (r^2 e ^{-\nu/2}
                   \xi)\Big\vert_{r=R} \nonumber \\
           &=& 0. \label{delp}
\end{eqnarray}
It is important to note that $\xi$ is finite, when $p$ vanishes at $r=R$.
The pulsations equation (\ref{puls}) is a Sturm - Liouville eigenvalue 
equation for $\sigma^2$, subject to the boundary conditions Eq.
(\ref{xi0}) and (\ref{delp}). As a consequence
the eigenvalues $\sigma^{2}$ are all real and  form an 
infinite discrete sequence $\sigma^2_o < \sigma^2_1 <\dots 
< \sigma^2_n <\dots\dots$, with the corresponding eigenfunction 
$\xi_0(r),~\xi_1(r),~...,\xi_n(r)$, where $\xi_n(r)$ has $n$ nodes.
It immediately follows that if the fundamental 
radial mode of a star is stable ($\sigma_o^2 > 0$), then all the 
radial modes are stable.
\par
We note that Eqs.(\ref{f}-\ref{a}) depend on the pressure and density profiles,
as well as on the metric functions  $\lambda(r),~\nu(r)$ 
of the nonrotating star configuration.
Those profiles are obtained by solving the Oppenheimer-Volkoff equations 
of hydrostatic equilibrium \citep{misner70}
\begin{eqnarray}
\frac {dp}{dr}&=& - \frac{G (\rho + p/c^2) (m + 4\pi r^3 p/c^2)} {r^2
                 (1-2 Gm/rc^2)}, \label{dpdr} \\ 
\frac {dm}{dr}&=& 4\pi r^2\rho , \\ \label{dmdr}  
\frac{d\nu}{dr}&=& \frac{2G}{r^2c^2} \frac{(m + 4\pi r^3
                   P/c^2)}{(1-2Gm/rc^2)}, \\ \label{dnudr}
\lambda &=& -\ln (1-2Gm/rc^2). \label{lamb}
\end{eqnarray}

Eqs. (\ref{dpdr}) -- (\ref{lamb}) can be numerically integrated 
for a given equation of state p($\rho$) and given central density to 
obtain the radius $R$ and gravitational mass $M = m(R)$ of the star.
Therefore the basic input to solve the structure and pulsation equations 
is the equation of state, $p=p(\rho)$.
It has been seen \citep{burgio01} that structure parameters of 
neutron stars are mainly dominated by the equation of state at high densities,
specifically around the core.
Since the oscillation features are governed by structure profiles of neutron
stars, it is expected to possess marked sensitivity on the high density
equation of state as well.

In this calculation, we adopted the equation of states, that have been derived
from BHF theory in a non-relativistic limit \citep{baldo98,baldo00} 
and from RMF theory
in a relativistic limit \citep{ghosh95,sahu00} without quark core.
These are denoted by BHF and RMF respectively.
Also, we took the equation of states of both BHF and RMF theories with 
quark core, considering the Wood-Saxon like density dependent 
parameterization of bag constant \citep{burgio01} in the quark sector.
They are represented by BHF+MW and RMF+MW.
Another set of equation of states with quark phase were taken with constant 
bag parameters in quark sector \citep{datta92} along with BHF and RMF 
theories, where the value of $B$ = 110 MeV fm$^{-3}$ was chosen in order
to ensure the presence of quark matter in the core, 
for completeness. These are correspondingly labeled by BHF+M and RMF+M.
We employed these three sets of equation of states to calculate 
the oscillation period $P$ ($=2\pi/c\sigma$) versus gravitational mass $M$
(in units of solar mass $M_\odot$).
The results are shown in figures 1-2.
\par
In both figures, in the upper panel the oscillation periods in 
seconds are shown versus the total gravitational mass, while in the 
lower panel the total gravitational mass versus central density is displayed.
For the sake of comparison, we show also the results for 
neutron stars without quark core.
If we carefully examine figure 1, we notice that the period of 
oscillations in BHF+MW and RMF+MW displays a small kink around the point 
where the mixed phases start, in the primary as well as in the higher modes.
In other words, the period increases and then decreases with respect 
to the usually decreasing trend observed in BHF and RMF.
This happens within a small 
range of gravitational masses, $0.4<M/M_\odot<0.7$, which corresponds 
to central density, $0.2<\rho_c/(10^{15} g ~cm^{-3})<1.7$, 
where the mixed phase regions are located.
The periods are slightly smaller than 0.4 ms  in BHF+MW and RMF+MW
models compared with the periods in BHF and RMF models for the fundamental 
mode and the similar trend has been seen in higher modes also.
When we compare BHF+M and RMF+M models with BHF and RMF models in  
figure 2, we notice that there are small kinks in the period of
oscillation in the mixed phase regions with 
gravitational masses $0.5<M/M_\odot<1.2$ and corresponding central densities, 
$0.6<\rho_c/(10^{15} g ~cm^{-3})<1.5$.
But these kinks are not as prominent as seen in BHF+MW and RMF+MW models.  
This is due to the fact that the transition from hadron to quark phase
is smooth in the case of a constant bag parameter. 
However, the periods of oscillations within these models are comparable with 
BHF and RMF models for both primary (larger than 0.4 ms) and higher modes.
Thus significant kinks are observed in BHF+MW and RMF+MW realistic
models, because the bag constant is density dependent.
The fundamental mode of oscillation periods for neutron stars with
and without quark core are found to have the range 0.2 --0.6 ms, the
only difference being a significant kink in neutron stars which are
associated with quark phase.
The substantial difference is observed in the fundamental modes of the
period of oscillations ($\sim$ .22 ms ) in the neutron stars which are 
composed of quark cores with density dependence bag parameters (BHF+MW and
RMF+MW) from that 
of normal neutron stars without quark core (BHF and RMF), at the maximum 
gravitational mass limit (see figure 1).
For higher modes the periods are $\le$ 0.3 ms for all the cases.
As another interesting point, we notice that all neutron stars 
with quark core have maximum gravitational masses around 1.5$M_{\odot}$.
\par
In summary, we have presented a calculation of the period of oscillations 
of neutron stars
by using the radial pulsation equations of nonrotating neutron star, as 
given by Chandrasekhar in general relativity formalism.
To solve the radial pulsation equations, one needs a structure profile of 
nonrotating neutron stars by employing realistic equation of states.
The equation of state we used here were derived from the non-relativistic and
relativistic formalism with quark phase at higher densities.
Since the quark matter is not well established, we explored the parameters
of quark matter compatible with the heavy-ion experiments at the point of 
possible formation of quark-gluon plasma.
Then the equation of state were constructed by using Glendenning condition for
mechanical and chemical equilibrium as a function of baryon and
electron density at the mixed phase, comprising with hadron, mixed 
and quark phases.
The main conclusion of our work is that the period of oscillations 
shows some significant kink against gravitational mass, if one uses
a realistic equation of state associated with density dependent bag constant.
These type of kinks are not present in conventional neutron stars, 
constituted by only hadrons. These kinks can be considered a distinct signature
of the quark matter onset in neutron stars.
%

\clearpage

\begin{figure}
\plotone{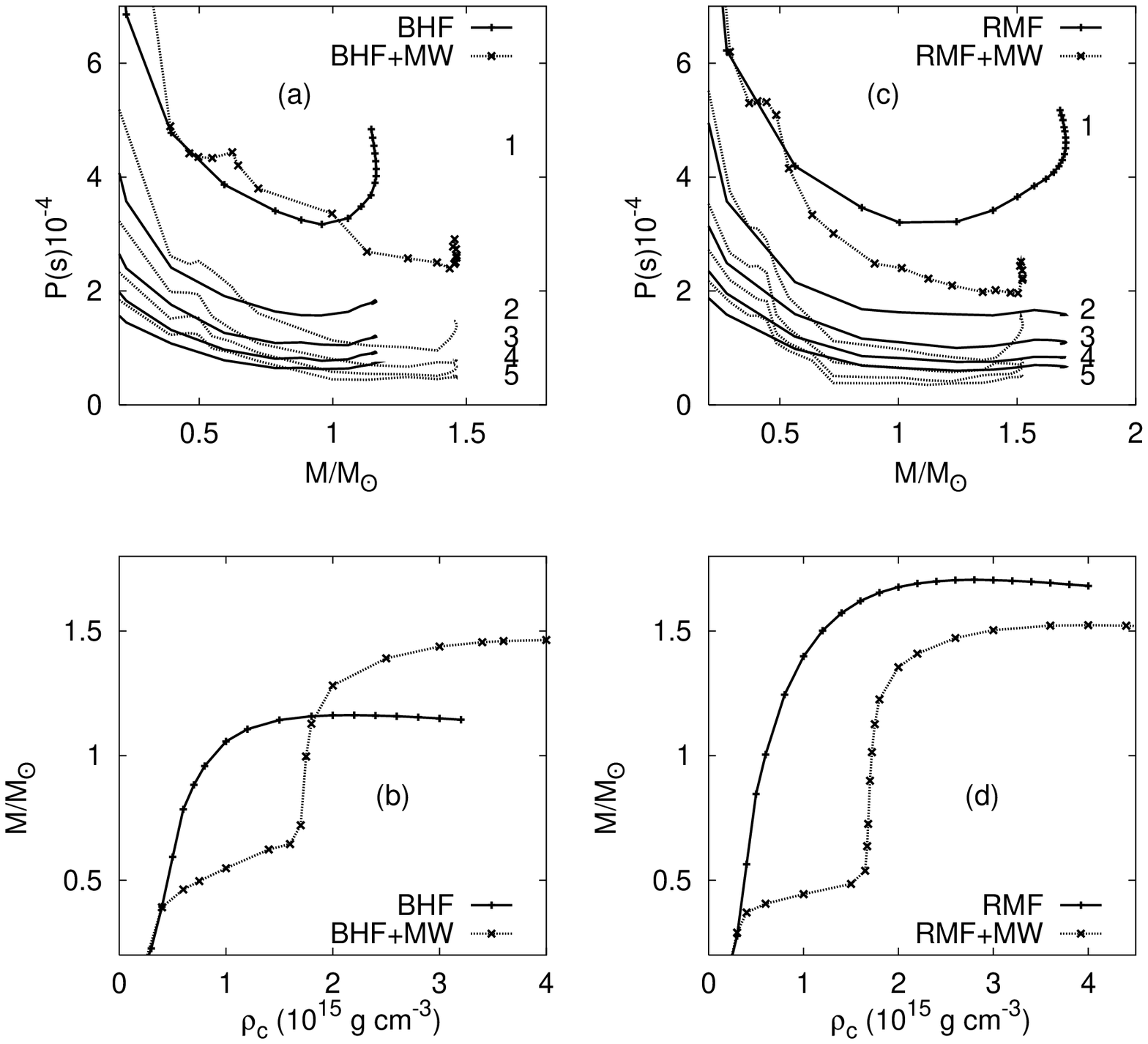}
\caption{In the upper panels the oscillation period in seconds is displayed 
vs. the gravitational mass
in units of the solar mass, whereas in lower panels the gravitational mass
is shown vs. the central density. Panels (a) and (b) show results for BHF
calculation for purely hadronic (solid line) and mixed hadron-quark matter
(dotted line). The bag constant is assumed to be density dependent. 
Labels 1,2, etc. indicate the higher modes. 
Panels (c) and (d) correspond to RMF calculations.\label{fig1}}
\end{figure}

\clearpage 

\begin{figure}
\plotone{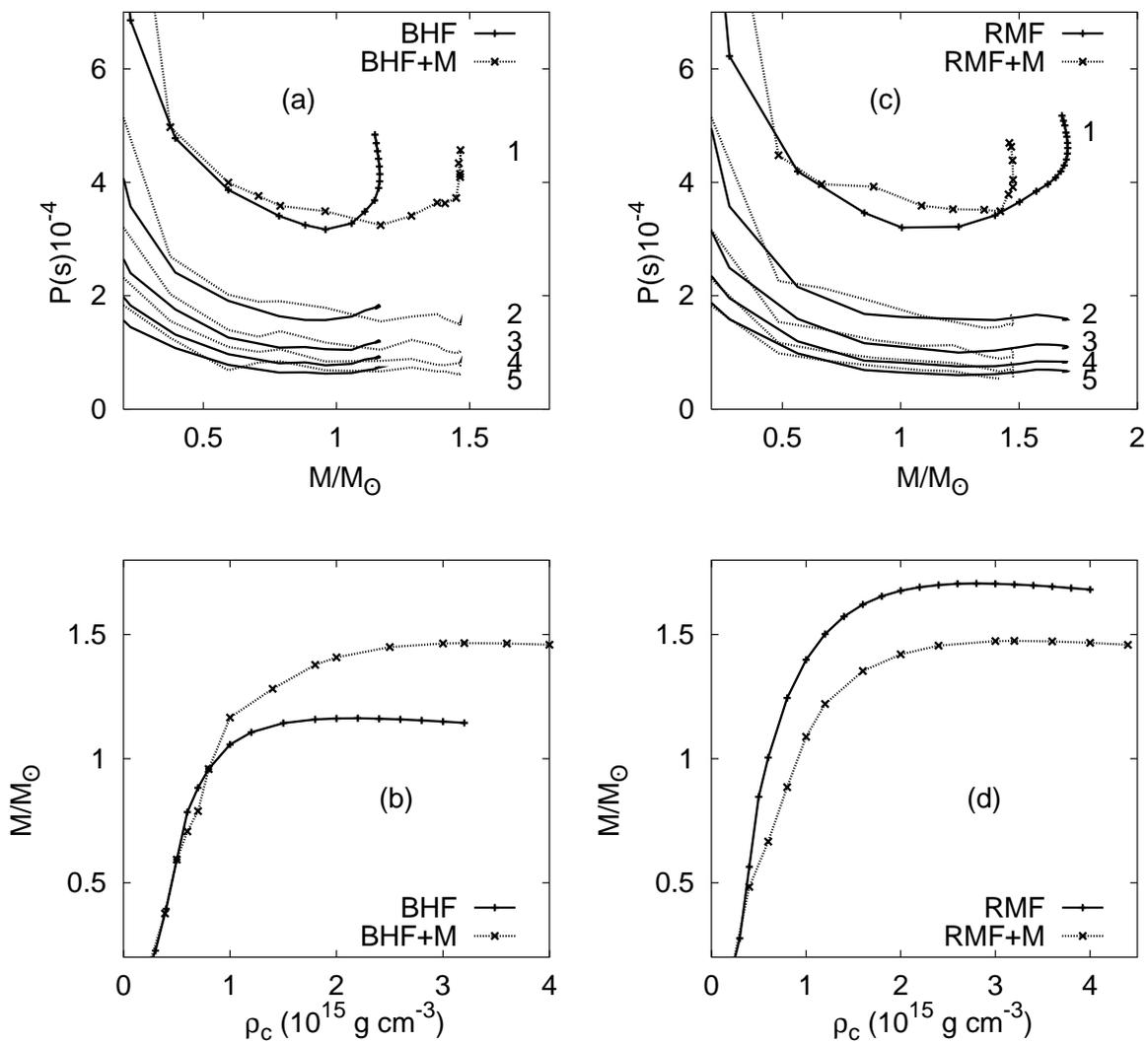} 
\caption{Same as figure 1, but for density independent bag constant.
See text for details.\label{fig2}}
\end{figure} 

\end{document}